\documentstyle[preprint,epsf,floats,aps] {revtex}

\begin{document}
\draft

\title{
Dynamical Quantum Processes of Molecular Beams at Surfaces:\\
Dissociative Adsorption of Hydrogen on Metal Surfaces}

\author{Axel Gross}

\address{Fritz-Haber-Institut der Max-Planck-Gesellschaft, Faradayweg 4-6, 
D-14195 Berlin-Dahlem, Germany
}

\maketitle

\begin{abstract}

Due to the improvement of computer power and the development of efficient
algorithms it is now possible to combine high-dimensional quantum dynamical 
calculations of the dissociative adsorption of molecular beams with 
reliable {\it ab-initio} potential energy surfaces (PES). 
In this brief review two recent examples of such studies of the systems
H$_2$/Cu(111), where adsorption is hindered by a noticeable energy barrier, 
and 
H$_2$/Pd(100), where activated as well as non-activated paths to adsorption 
exist, are presented. The effect of lateral surface corrugations on 
the sticking probability in the tunneling and the classical regime and 
the role of additional parallel momentum are discussed in the
context of the H$_2$/Cu(111) results. For the system H$_2$/Pd(100) it 
is shown that the initial decrease of the sticking probability with
increasing kinetic energy, which is usually attributed to a precursor 
mechanism, can be explained by dynamical steering. In addition, the influence 
of rotation on the adsorption and desorption dynamics is examined.

\end{abstract}


\section{Introduction}

In recent years the dynamics of dissociative adsorption and associative
desorption of hydrogen has been studied in great detail, 
experimentally as well as theoretically 
(see, e.g., Refs.~\cite{Ren94} and \cite{Hol94}).
One of the key motivation for these investigations is that by studying
this simple surface reaction one hopes to learn about the crucial
principles underlying more complex surface reactions which occur, e.g.,
in heterogenous catalysis. In the following I will focus on 
quantum dynamical investigations aimed at the understanding of dissociation.
For a much more detailed discussion see the recent excellent review 
by Darling and Holloway \cite{Dar95}.

Chemical reactions at surfaces represent
the interaction of a system with a limited number of degrees of freedom 
(the molecule) with the half-infinite solid which posesses in principle 
an infinite number of degrees of freedom. However, for the dissociative
adsorption of hydrogen, especially on metal surfaces, there is little
energy transfer to the substrate phonons due to the large mass mismatch.
The crucial process in the understanding of the dissociation is the 
conversion of the kinetic and internal energy of the molecule into 
translational energy of the atoms relative to each other, the subsequent 
energy dissipation to the solid is not relevant for the dissociation.
In addition, electronic levels of hydrogen molecules in front of metal 
surfaces are broadened significantly \cite{Ham94} so that electronic
excitations are very effectively quenched. 

This situation seems to be different for hydrogen adsorption on 
semi-conductor surfaces,
especially in the well studied system H$_2$/Si. It had been proposed 
\cite{Bre94a} that the experimental puzzle of low sticking coefficient 
and almost thermal distribution in desorption - the former result indicating 
a high barrier to adsorption, the latter a low barrier 
(see \cite{Kol94} and references therein) - can be reconciled if lattice
distortions on adsorption are taken into account. The strong rearrangement of 
the silicon surface due to hydrogen adsorption has been confirmed in
total-energy calculations \cite{Kra94,Peh95}, and also the predicted
strong surface temperature dependence of the sticking probability of
H$_2$/Si \cite{Bre94a} has been found experimentally \cite{Kol94,Bra95}.

The dynamics of dissociative adsorption of hydrogen on the more densely 
packed metal surfaces, however, can be described using low-dimensional 
potential energy surfaces (PES) because of the large mass mismatch, the
smaller space for surface rearrangements,
and the quenched electronic excitations \cite{Dar95}.
In principle the PES should be six-dimensional describing all the
degrees of freedom of the hydrogen molecule. However, up to recently there 
had been no six-dimensional quantum dynamical investigations of the adsorption
and desorption dynamics because the calculations were computationally too 
demanding. Hence the quantum dynamicists were restricted to low-dimensional
model calculations. Since the crucial coordinate for the dissociation
process is the intramolecular spacing, a large number of studies were
devoted to the dynamics on two-dimensional so-called "elbow" potentials
where the H$_2$ center of mass distance from the surface and the H-H
distance were considered \cite{Jac87,Hal90,Dar92a,Kue91,Sch92,Dar92b}.  
These studies provided us with a qualitative understanding of the
topological features that realistic potentials should have 
in these two coordinates in order to
describe surface reactions. Further quantum mechanical model studies
coupled the two-dimensional elbow potential to one surface oscillator
in order to account for recoil and surface temperature effects
\cite{Bre94a,Han90,Gro94e,Dar94a}. Just recently there has been a large number
of papers dealing with the effects of rotations on the sticking
probability, especially with steric effects and the non-monotonous
dependence of the sticking probability on rotational quantum number~$j$
in the system H$_2$/Cu(111) \cite{Mow93,Dar94b,Bru94,Kum94,Dai95}.

Although already in the earliest studies on dissociative adsorption 
\cite{Len32}
the important role of surface corrugation was stressed, except for in the
studies of N{\o}rskov and co-workers \cite{Nor81,Kar87,Eng92} and Holloway
and co-worders \cite{Hal88,Dar90,Dar92c}, little attention  has been paid to 
the influence of corrugation on the adsorption dynamics. There are two
main reasons for this. Firstly a large number of adsorption systems show the
so-called normal-energy scaling \cite{Ren89,Mic91,Ret92}, 
i.e., the sticking probability
is a function of the normal component of the kinetic energy of the 
impinging molecules alone. This scaling behaviour is usually associated
with a flat, structureless surface. And secondly, the magnitude of the
corrugation had to be more or less guessed since the mapping
out of the whole PES by {\em ab initio} methods was much too time-consuming.
Since also the electronic density in front of metal surfaces is rather
smeared out \cite{Che75}, the picture of low-corrugated metal surfaces was 
widely accepted.

Due to the improvement of computer power and the development of 
efficient algorithms it just recently has become possible to map out
the six-dimensional PES of hydrogen dissociation on metal surfaces
\cite{Ham94,Whi94,Wil95,Wil95b} by reliable density-functional calculations.
It came as a surprise when elaborate studies of 
the H$_2$/Cu system \cite{Ham94,Whi94} revealed that the PES is strongly
corrugated although this system obeyed normal energy scaling in the beam
experiments \cite{Mic91,Ret92}. These theoretical findings renewed the
interest of the role of corrugation for the dissociation dynamics
\cite{Dar94,Gro95a,Bre95}. 
In a three-dimensional model study Darling and Holloway \cite{Dar94}
showed how the apparent contradiction between normal energy scaling
and strongly corrugated PES can be reconciled if the PES exhibits
topological features now termed "balanced corrugation" \cite{Dar95}.

The availability of high-dimensional potential energy surfaces 
has caused new efforts for the improvement of the quantum dynamical
algorithms. While up to recently it was believed that the inclusion of all
molecular degrees of freedom in a quantum dynamical calculation is
too computationally expensive \cite{Kum94,Dai95}, there is now the first
six-dimensional quantum study of the dissociative adsorption and
associative desorption \cite{Gro95b}. High-dimensional quantum dynamical 
studies using {\em ab initio} potentials serve a twofold purpose. Firstly they
can yield a quantitative as well as also a novel qualitative understanding of 
the dissociation process because all relevant degrees of freedom are included.
But secondly also the reliabiblity of the total-energy calculations can be 
checked. The results for the H$_2$/Cu system showed that one has to 
go beyond the local-density approximation (LDA) in the treatment of 
exchange-correlation effects in order to obtain realistic barrier heights
\cite{Ham94,Whi94}. But still the non-local corrections are approximative
and their validity is highly debated, especially in the H$_2$/Si(100)
system \cite{Kra94,Peh95,Nac95}. Since from just looking at a PES one cannot
extract whether the PES is accurate, one has to perform dynamical calculations
{\em on} this PES in order to find out. 

In the following I will present two examples of high-dimensional quantum
dynamical studies on {\em ab initio} potential energy surfaces. The
first example deals with the H$_2$/Cu(111) system \cite{Gro94b} where 
dissociation is hindered by noticable barriers. I will focus on the effect 
of lateral surface corrugation on the sticking probability in the tunneling 
and the classical regime and on the role of additional parallel momentum.
The second example is then devoted to H$_2$/Pd(100) \cite{Gro95b}, a system
where activated as well as non-activated paths to dissociative adsorption
exist. It will be shown that the initial decrease of the sticking probability
with kinetic energy found in this system \cite{Ren89} is not due to a
precursor mechanism, as was commonly believed, but can be explained by
dynamical steering. In addition, rotational effects will be discussed.
The paper ends with some concluding remarks.

\section{H$_2$/C\lowercase{u}(111)}

H$_2$/Cu has been the benchmark system for the study of dissociative
adsorption. An abundance of experimental \cite{Ret92,Ret95,Hay89,Ang89}
and theoretical \cite{Dar92a,Kue91,Mow93,Dar94b,Bru94,Kum94,Dai95}
investigations about this system exists, just to mention a few. The
calculated barrier heights varied substantially over the years (see
ref.~\cite{Ham94}) and the discussion still goes on \cite{Whi94,Wie95}.
For the H$_2$/Cu(111) it was found that the barrier to dissociative
adsorption for the molecular axis parallel to the surface varied between
0.73 and 1.43~eV within the surface unit cell for a four-layer substrate
in the slab calculations. Convergence tests suggested that the minimum
barrier should be lowered to 0.5~eV. A five-dimensional
parametrisation of the {\em ab initio} PES has served as an input
for a dynamical study of the adsorption and desorption \cite{Gro94b},
were the consistent four-layer results have been used. 
Further convergence tests which were finished after the
completion of the dynamical calculations showed that the whole
barrier region of the PES should be shifted by approximately the 
same amount as the minimum barrier \cite{Ham95}.
In the dynamical study the three center-of-mass coordinates and the 
interatomic
spacing of the H$_2$ molecule have been treated quantum mechanically, while
the azimuthal orientation of the molecule has been taken into account in a
classical sudden approximation which works quite well in the H$_2$/Cu
system \cite{Gru93}. The polar orientation is not varied which means that
the molecular axis is kept parallel to the surface.
The quantum dynamics were determined in a
time-independent coupled-channel scheme using the concept of the
{\em local reflection matrix} \cite{Bre93,Chi94} and the {\em inverse
local transmission matrix} \cite{Bre94}. This very stable method
also yields the wave function of the scattering problem \cite{Bre95}.
It is closely related to the logarithmic derivative of the solution matrix
and thus avoids exponentially increasing evanescent waves which 
cause numerical instabilities.

\begin{figure}[ht]
\unitlength1cm
\begin{center}
   \begin{picture}(10,6)
      \includegraphics{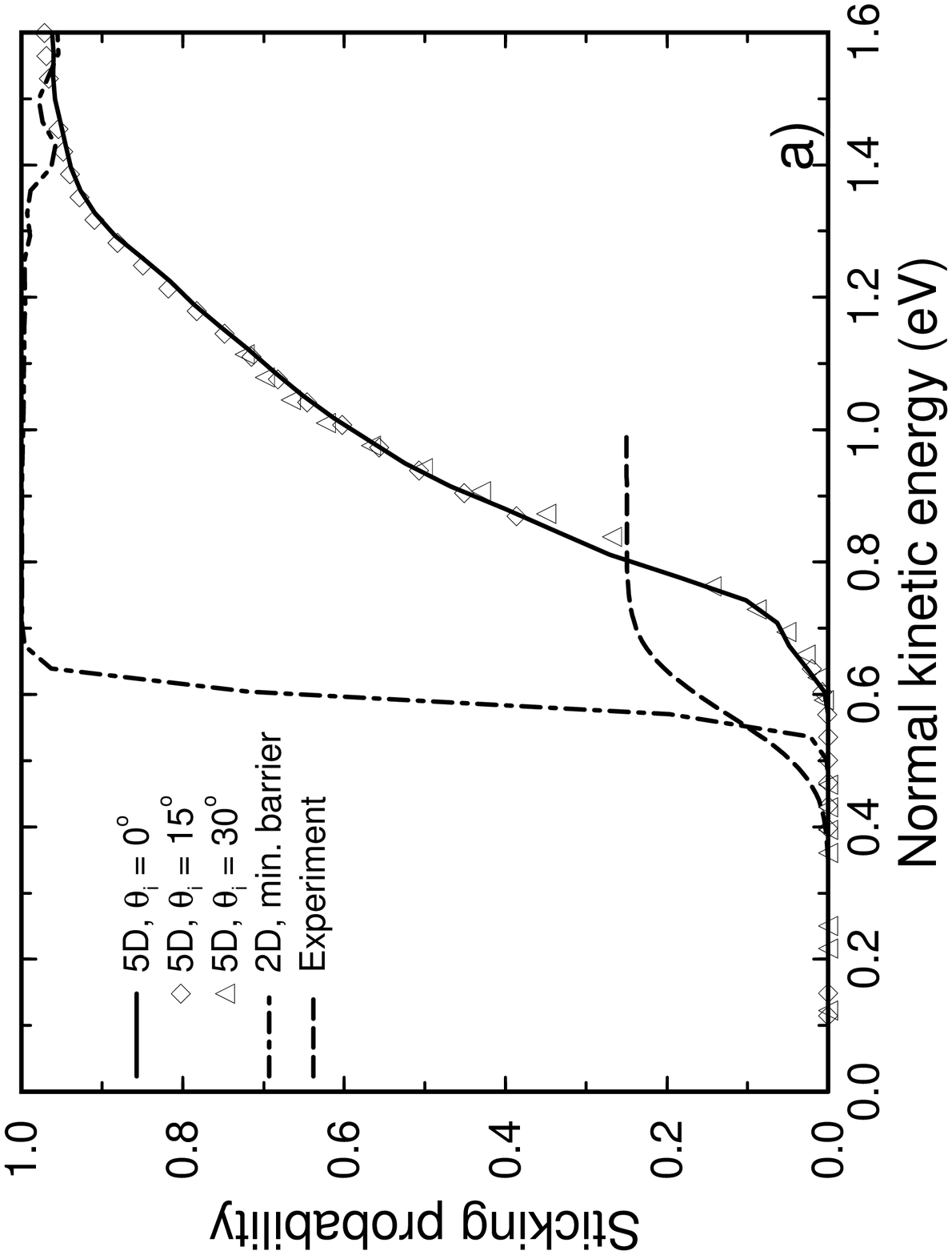}
   \end{picture}

   \begin{picture}(10,6)
      \includegraphics{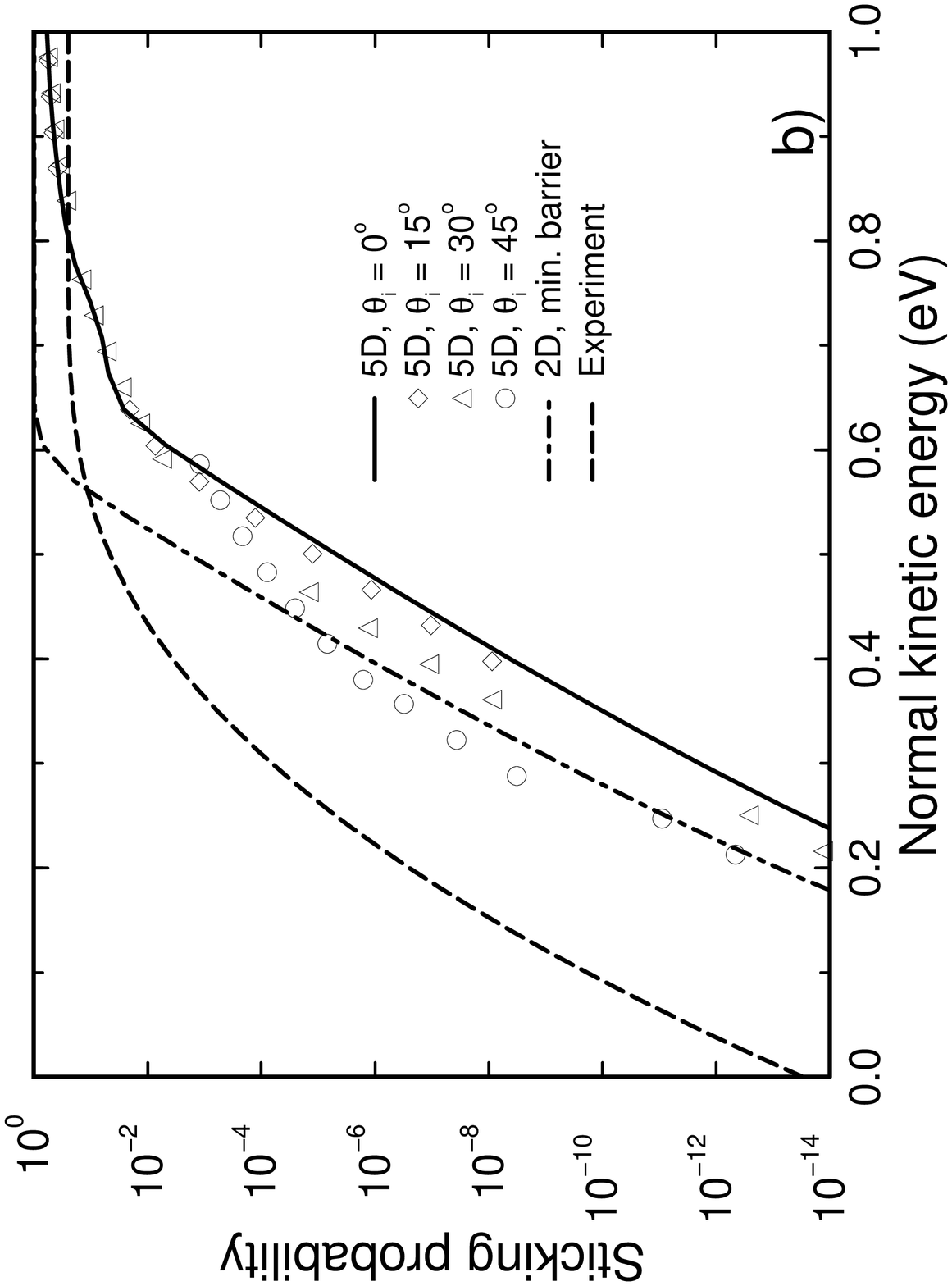}
   \end{picture}
\end{center}
   \caption{Sticking probability versus normal kinetic energy for
molecules initially in the vibrational ground state. 
a) Linear plot, b) logarithmic plot (note the different energy range). 
5D-calculations for different incident angles at the corrugated
surface: solid line $\theta_i = 0^{\circ}$, 
$\Diamond \ \theta_i = 15^{\circ}$, $\bigtriangleup \ \theta_i = 30^{\circ}$,
$\bigcirc \ \theta_i = 45^{\circ}$;
Dash-dotted line: 2D-calculations corresponding to a flat surface 
with the minimum barrier (from ref.~\protect{\cite{Gro94b}}).
Dashed line: Experimental curve (from ref.~\protect{\cite{Ret95}}).} 
\label{stickh2cu}
\end{figure}

The results of the quantum dynamical calculations of the sticking probability
of H$_2$/Cu(111) for molecules initially in the vibrational ground state 
under normal and non-normal incidence are shown in fig.~\ref{stickh2cu} 
versus the normal kinetic energy in a linear and in a logarithmic plot.
For comparison also the results of two-dimensional calculations with the
minimum barrier are shown. I will first focus on the normal-incidence data 
in the low-energy regime (fig.~\ref{stickh2cu}(b)) where sticking 
is impossible classically. Comparing the 2D and the 5D results shows 
that the effect of corrugation in the tunneling regime is an almost 
energy-independent suppression of the sticking probability, in this case 
by about two orders of magnitude. Since the increase of the sticking 
probability in 2D calculations is determined by the barrier 
width \cite{Dar92a}, this indicates that at energies below the minimum 
barrier height the adsorption dynamics are governed by the minimum barrier 
path. This result is also confirmed by wave function plots \cite{Gro95a}.
It can be simply understood by the fact that the transmission through a 
barrier
in the tunneling regime is exponentially suppressed which on the other
hand leads to an exponential preference of the minimum barrier path compared
to all other paths.

In the energy regime where sticking classically is possible (which will be
referred to as the "classical regime" in the following) the sticking curves
of the 2D and the 5D calculations are no longer parallel 
(fig.~\ref{stickh2cu}(a)). Indeed the sticking probability in the 5D
calculations starts rising in the linear plot at an energy corresponding
to the minimum barrier height and reaches unity at an energy 
which corresponds to the maximum barrier height 
(the onset of the sticking in the linear plot occurs already at approximately 
0.15~eV below the minimum barrier height due to the softening of the 
H$_2$~bond at the surface and the accompanying reduction of the zero-point 
vibrational energy). Obviously the whole distribution of the barrier heights
determines the sticking probability. At these high kinetic energies (larger 
than 0.6~eV) the dynamics in the classical regime can be described in the 
sudden limit. A molecule impinging on the surface sticks or not according to
whether its energy is larger than the barrier it hits or not, there are
little reorientation and steering effects. This means
that the sticking in this energy regime can be understood  in terms of the
available phase space for dissociation which is the basic assumption
underlying the so-called hole-model \cite{Kar87}.

In the first presentation \cite{Gro94b} the theoretical results were compared
to data compiled by Michelsen and Auerbach \cite{Mic91}. They fitted
existing adsorption data and, via the principle of detailed balance,
also desorption data to a functional form of the sticking probability
of the form  
\begin{equation}
S_0 (E) \ = \ \frac{A}{2} \ \left[ 1 + \tanh \left(
\frac{E - E_0}{W}\right) \right].
\end{equation}
These experimental data were at variance with the theoretical data of
ref.~\cite{Gro94b} especially
at low and at high energies. In the meantime a new, very detailed 
experimental study of the adsorption and desorption of H$_2$/Cu(111)
has been published by Rettner {\it et al.} \cite{Ret95}. To describe
the sticking function they used
\begin{equation}
S_0 (E) \ = \ \frac{A}{2} \ \left[ 1 + \mbox{erf} \left(
\frac{E - E_0}{W}\right) \right],
\end{equation}
because this functional form reproduces the adsorption data better
at low energies \cite{Mic93}. And indeed,
while in the old parametrisation the experimental sticking probability
did not drop below 10$^{-4}$ \cite{Mic91}, now at low energies
in fig.~\ref{stickh2cu}(b) the experimental and the theoretical
curve no longer disagree substantially. The two curves are almost
parallel, whereby the shift of  approximately 0.25~eV can be
related to the minimum barrier height in the calculations which seems
to be to too high by the same amount \cite{Ham94}. 
This indicates that, except for the barrier height, 
the minimum barrier path is properly described by
the {\em ab initio} PES. The same shift of 0.25~eV 
between theory and experiment is also seen at the onset of sticking
in the linear plot, fig.~\ref{stickh2cu}(a) (note the different energy scale
in the linear and the logarithmic plot). However, while the
experimental curve saturates at a sticking probability of 0.25, the
theoretical curve reaches almost one. The issue of the maximum
sticking probability for H$_2$/Cu(111) is still strongly debated.
Rettner {\it et al.} \cite{Ret95} state that their experimental sticking 
probabilities could equally well fitted with A values between 0.15 and
0.5 since the measurements do not extend to energies above 0.55~eV.
Just one fitting parameter $W$ for the width of the sticking curve
may not be sufficient, since the increase of the sticking probability is
determined by two different properties: in the tunneling regime by
the barrier width, and in the classical regime by the distribution
of the barrier heights. On the other hand the polar rotation of the
molecule had not been considered for the theoretical results in
fig.~\ref{stickh2cu}. Particles that hit the surface in the
upright orientation can not dissociate. So the inclusion of 
the polar orientation will probabably reduce the maximum sticking 
probability since at high energies 
molecules will not have enough time to reorientate towards a more
favorable configuration towards dissociative adsorption during the
scattering process.

\begin{figure}[thb]
\unitlength1cm
\begin{center}
   \begin{picture}(10,6)
      \includegraphics{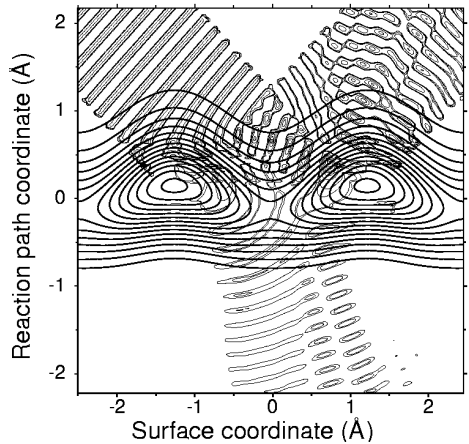}
   \end{picture}
\end{center}
   \caption{Positive real part of the wave function of H$_2$ scattered
            at a PES with one-dimensional lateral corrugation simulating
            H$_2$/Cu(111). The incident beam has a kinetic energy of 
           $E_i = 1.08$~eV, the incident angle is $\theta_i = 45^{\circ}$.
          The contour spacing for the potential (thick lines) is 0.1~eV.
          The positions of the Cu atoms are at the potential maxima 
          (from ref.~\protect{\cite{Gro94b}}).}
\label{h2cuwfpes}
\end{figure}

With regard to the effect of additional parallel momentum,
fig.~\ref{stickh2cu}a) shows that for non-normal incidence 
the sticking probabilities fall upon the normal incidence sticking
curve if they are plotted versus the normal kinetic energy, at least
in the classical regime. This normal-energy scaling is surprising 
considering the strong corrugation of the surface which is shown
in fig.~\ref{h2cuwfpes}. Darling and Holloway \cite{Dar94} have first 
addressed this issue. They showed that the variation of
the barrier height (energetic corrugation) and the variation of the
barrier position (geometric corrugation) have opposing effects as far as 
the role of additional parallel momentum is concerned. The plot of 
the wave function \cite{Bre95,Gro94b}
scattered at the corrugated PES in fig.~\ref{h2cuwfpes}
illustrates these effects. In order to make both the incoming and
the reflected part of the beam visible in one single plot, the
lateral extension of the beam perpendicular to its propagation
direction, which is in principle infinite, is restricted. Thus by
following the wave fronts the propagation of the beam and its splitting 
in a reflected and transmitted part can be followed. 
Due to the energetic corrugation molecules 
with additional parallel momentum will sample a range of energy barriers 
which leads to a averaging process over the barriers. 
If their normal kinetic energy is less than the mean barrier height then 
this averaging process leads to a suppression of the sticking probability 
\cite{Gro95a}. Fig.~\ref{h2cuwfpes} shows that molecules which encounter 
the high-barrier sites are scattered back into the gas phase.

On the other hand the area in the surface unit cell where the 
propagation direction of the incoming beam is aligned to the
local gradient of the potential becomes actually larger for additional
parallel momentum due to the geometric corrugation. This results
in an enhancement of the sticking probability \cite{Dar94,Gro95a}. Now
if the maximum barriers are at a greater distance from the surface,
the opposing effects of energetic and geometric corrugation can
balance \cite{Dar95,Dar94} leading to normal-energy scaling in spite of
the strong corrugation.

The question now arises whether this is just a coincidence that 
nature has hidden the fact of the strongly corrugated surfaces by 
these peculiar balancing features of the PES. However, total energy
calculations show that usually the larger barriers to dissociative 
adsorption are in front of on-top positions \cite{Ham94,Whi94}.
And there the interaction between molecule and surface occurs further
away from the surface as compared to, e.g., bridge and hollow sites
\cite{Whi94,Gro94b} since in front of the atoms the electron density
spills further out into the vacuum. 
So the topological features responsible for the normal-energy
scaling do not seem too peculiar regarding the underlying principles
of the chemical interaction.

Figure~\ref{stickh2cu}(b) also reveals that in the tunneling regime
additional parallel momentum enhances sticking. This is due to the fact
that in the realm of tunneling dynamics parallel momentum is effectively
converted into normal momentum for energetic as well as for geometric
corrugation \cite{Gro95a}. The observation that in beam experiments normal
energy scaling is found for all energies in the system H$_2$/Cu(111)
\cite{Ret92,Ret95} can be 
explained by the fact that a low kinetic energies sticking in the beam
experiments is dominated by the vibrationally excited molecules, and for
these molecules the range where normal-energy scaling is obeyed is shifted
to lower energies \cite{Bre95}.

\section{H$_2$/P\lowercase{d}(100)}

Molecular beam experiments of the dissociative adsorption of H$_2$ on various
transition metal surfaces like Pd(100) \cite{Ren89}, Pd(111) and Pd(110)
\cite{Res94}, W(111) \cite{Ber92}, W(100) \cite{Ber92,But94,Aln89}, 
W(100)-c(2$\times$2)Cu \cite{But95} and Pt(100) \cite{Dix94}  show
that the sticking probability initially decreases with increasing kinetic 
energy
of the beam in these systems. Such a behaviour is usually attributed to a
precursor mechanism: before dissociation the molecules are temporarily trapped
in a molecular adsorption state, the so-called precursor state, where they
accomodate to the surface temperature. In order to be trapped, the molecules
have to lose their initial kinetic energy which is assumed to happen primarily
by energy transfer to the substrate phonons \cite{Ren94}. And it is this
trapping probability which decreases with increasing energy and thus determines
the sticking probability for dissociative adsorption. At higher energies
the sticking probability rises again in all systems mentioned above indicating
that the adsorption via direct activated paths becomes dominant.

However, for hydrogen adsorption the large mass mismatch between adsorbate
and substrate should make the energy transfer process inefficient. Together
with the missing surface temperature dependence this led to a dynamical 
steering
process being proposed in order to explain the initial decrease of the sticking
probability \cite{Aln89,Dix94}, but there had been no theoretical confirmation 
whether this mechanism could be effective. Recent density-functional
theory calculations show that there exist non-activated as well as activated
paths to dissociative adsorption in the system H$_2$/Pd(100), but no
molecular adsorption well \cite{Wil95}. And three-dimensional quantum dynamical
calculations using a model PES with such features suggested that a steering
mechanism can indeed be responsible for the initial decrease of the
sticking probability \cite{Gro95a}, but large quantitative discrepancies to
the experiment remained.

\begin{figure}[htb]
\unitlength1cm
\begin{center}
   \begin{picture}(10,6.0)
      \includegraphics{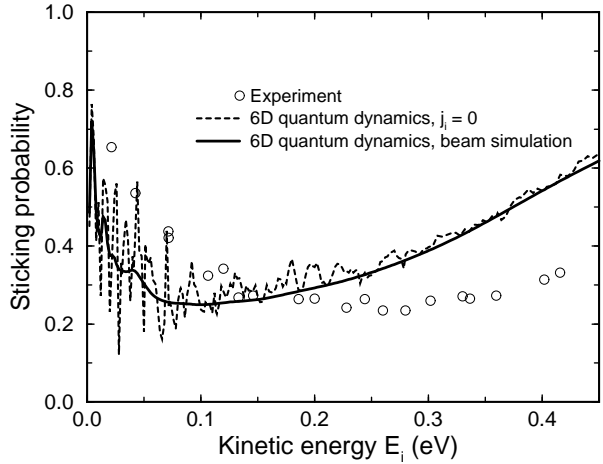}
   \end{picture}

\end{center}
   \caption{Sticking probability versus kinetic energy for
a H$_2$ beam under normal incidence on a Pd(100) surface.
Experiment: circles (from ref.~\protect{\cite{Ren89}}); theory:
H$_2$ molecules initially in the rotational ground state (dashed line)
and with an initial rotational and energy distribution 
adequate for molecular beam experiments (solid line)
(from ref.~\protect{\cite{Gro95b}}). }

\label{h2pdstick}
\end{figure}

Using a parametrization of the {\it ab initio} PES of H$_2$/Pd(100) 
dynamical calculations of the dissociative adsorption and associative
desorption have been performed where for the first time {\em all} six
degrees of freedom of the hydrogen molecule have been treated quantum
dynamically \cite{Gro95b}. Figure~\ref{h2pdstick} shows a comparison
between experiment \cite{Ren89} and the 6D calculations. The dashed curve
which corresponds to molecules initially in the rotational ground state
exhibits a strong oscillatory structure which is at variance with the 
experimental data. Performing three-dimensional
model calculations on a PES with similiar features as the {\em ab initio}
PES of H$_2$/Pd(100) it has been shown that the quantum nature of the hydrogen 
beam leads to an oscillatory structure in the 
sticking probability\cite{Gro95a}.
Whenever the kinetic energy becomes large enough to open up new scattering 
channels, the reflection rate has a maximum. Since the reflection rate and
the sticking probability are related by unitarity, this means that at these
energies the sticking probability shows a minimum. 
Strong oscillations in the sticking probability were also found 
in three-dimensional wave-packet calculations \cite{Dar90} where the PES 
included a precursor well. They were attributed to the existence of resonance 
states in the attractive well which, however, does not exist in the 
{\em ab inito} potential of H$_2$/Pd(100) \cite{Wil95,Wil95b}.

In order to compare the results of the 6D
calculations with the beam experiment data one has
take into account that the experimental beam does not correspond to a
single quantum state. Firstly, excited rotational states are populated
according to a Boltzmann distribution with a temperature of 0.8 of the
nozzle temperature in the case of hydrogen beams \cite{Ren89}.
And secondly, the beam is not strictly monoenergetic, 
but has a certain energy spread of typically 
$\Delta E / E_i = 2 \Delta v / v_i = 0.2$ \cite{Ren89} ($E_i$ and $v_i$ are
the initial kinetic energy and velocity, respectively).
If the theoretical results are averaged over initial states according
to the experimental conditions (solid line in fig.~\ref{h2pdstick}), the
strong oscillations disappear.

The resulting theoretical curve agrees quite satisfactory with experiment.
Although no precursor state exists in the PES and the energy transfer to 
substrate phonons is not taken into account due to the fixed substrate
atoms, the initial decrease of the sticking probability with increasing
kinetic energy is well reproduced. Since the precursor mechanism
cannot be responsible, only a dynamical steering effect can cause the initial
decrease. This steering effect is confirmed in the coupled-channel 
calculations
by the fact that at low energies more channels are needed in order to get
converged results than at high energies. Usually it is the other way around
since at higher energies more channels are energetically accessible.
This unusual behavior indicates that at low energies there is a strong
dynamical redistribution among the different channels while at high
energies the dynamics is closer to the adibatic limit with little
transitions between the channels.

\begin{figure}[htb]
\unitlength1cm
\begin{center}
   \begin{picture}(10,6.0)
      \includegraphics{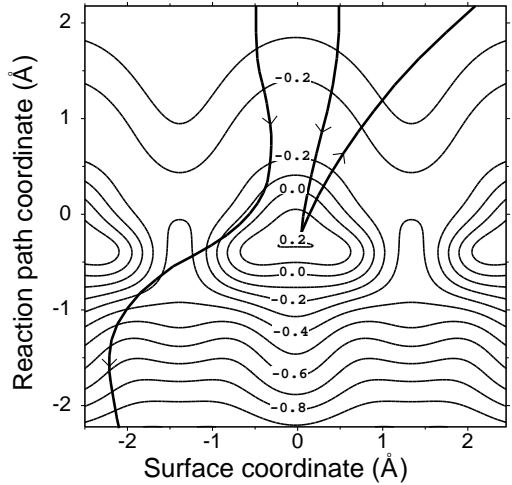}
   \end{picture}

\end{center}
   \caption{Illustration of the steering effect. The contour
lines show the potential energy surface in eV along a two-dimensional cut 
through the six-dimensional configuration space of a hydrogen molecule
with the molecular axis is parallel to the surface. 
The Pd atoms are at the potential maxima. 
The surface coordinate corresponds to the $X$-axis, while the reaction path
coordinate $s$ for $s \rightarrow \infty$ is equivalent to the H$_2$ center 
of mass distance from the surface, for $s \le - 2.5$~{\AA} to the distance 
between two adsorbed H~atoms.
The left trajectory corresponds to a slow molecule (kinetic energy 
$E_{kin} = 0.05$~eV), the right trajectory to a fast molecule 
($E_{kin} = 0.15$~eV)(from ref.~\protect{\cite{Gro95b}}).}

\label{h2pdtraj}
\end{figure}

The classical analogue to the quantum dynamical description of the steering
effect is illustrated in fig.~\ref{h2pdtraj} which shows 
a two-dimensional cut through the six-dimensional PES together 
with two typical trajectories of impinging H$_2$ molecules.
By symmetry, both trajectories have equivalent impact parameters,
but while the slower molecule (left trajectory) can be steered towards
a non-activated path to adsorption by the attractive forces, the other 
molecule (right trajectory) is too fast for the
forces to divert it significantly. It hits the repulsive part of 
the potential and is reflected back into the gas phase. 
By further increasing the kinetic energy the molecule will eventually
have enough energy to directly cross the barrier which leads to the increase 
of the sticking probability at higher energies (see fig.~\ref{h2pdstick}).

\begin{figure}[htb]
\unitlength1cm
\begin{center}
   \begin{picture}(10,6.0)
      \includegraphics{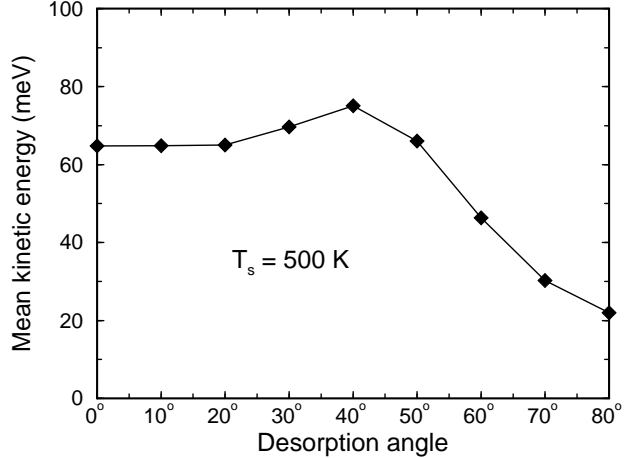}
   \end{picture}
\end{center}
   \caption{Angular distribution of the mean kinetic energy 
           of molecules desorbing from a one-dimensional corrugated
           surface simulating H$_2$/Pd. Surface temperature $T_s = 500$~K 
           (from ref.~\protect{\cite{Gro95a}}).}
   \label{wkinpd}
\end{figure}

The steering effect should also be observed in desorption properties, e.g.,
the mean kinetic energy in desorption. In ref.~\cite{Gro95a} three-dimensional
dynamical calculations on a model potential with non-activated as well as 
activated paths to dissociative adsorption have been performed where the
PES looked quite similiar to fig.~\ref{h2pdtraj}. In order to desorb the
molecules have to have enough kinetic energy normal to the surface to
overcome the barrier for desorption. This corresponds to trajectories
starting from below in fig.~\ref{h2pdtraj}. Now most desorbing molecules 
propagate along minimum barrier paths \cite{Gro95a}. Except for the molecules
following the high-symmetry path there are forces acting on the molecules
which will divert them from the normal desorption direction. And again,
the slower the molecules are, the more they will be diverted. This also
means that at high desorption angles primarily low-energy molecules will
be found. Indeed the 3D model calculations follow this trend 
(see fig.~\ref{wkinpd}) \cite{Gro95a}. While for normal desorption the mean 
kinetic energy is in equilibrium with the surface temperature (note that in 
two dimensions the flux-corrected mean kinetic energy of a gas in equilibrium 
is 3/2 kT), for high desorption angles the mean kinetic energy is strongly
lowered. This means that a peaked mean kinetic energy in desorption
is not always indicative of non-thermal processes \cite{Her95}. In addition,
it shows that one-dimensional models \cite{vWil68} are not sufficient in 
order to describe angular distributions in desorption.

A dynamical degree of freedom that has not been considered so far in this 
brief
review is the rotational degree of freedom. As for the influence of additional 
rotational motion on the sticking probability, there are two different
mechanisms operative. Just recently it has been pointed out \cite{Dar94b} 
that, apart from the change in moment of inertia, an isomorphism exists 
between surface corrugation and a planar rotor which means that there is
a close correspondence in the way the lateral surface corrugation and the
orientational anisotropy of the molecule-surface potential enter into
the dynamics. The barrier to dissociative
adsorption strongly depends on the molecular orientation \cite{Ham94,Wil95b}.
Thus additional rotational motion acts like additional parallel momentum
in the case of energetic corrugation. Due to the range of barrier heights
which is probed by the rotating molecule the sticking probability is
effectively suppressed \cite{Dar94b}. However, in the case of the H$_2$/Cu
system due to the late barrier to adsorption the molecular bond is strongly
extended when the barrier region is crossed. The resulting increase in the 
moment of inertia causes a decrease of the rotational energy. This leads
to a lowering of the effective barrier with increasing rotational quantum
number $j$ and to an enhancement in the sticking probability for high 
$j$~values \cite{Dar94b,Bru94} in agreement with the experiment \cite{Mic93}.

\begin{figure}[htb]
\unitlength1cm
\begin{center}
   \begin{picture}(10,6.0)
      \includegraphics{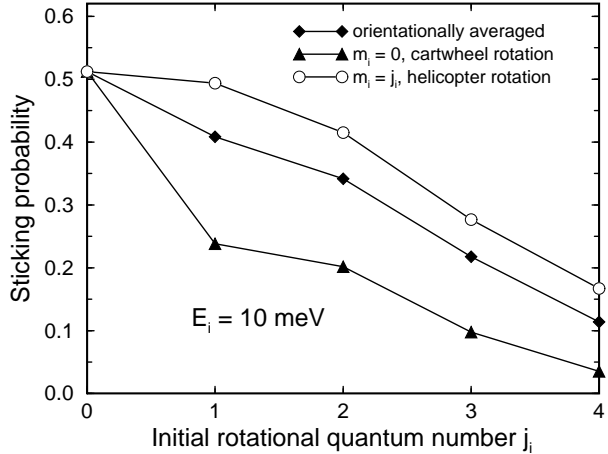}
   \end{picture}

\end{center}
   \caption{Sticking probability versus initial rotational quantum state $j_i$
   for the system H$_2$/Pd(100).
   Diamonds: orientationally averaged sticking probability 
   (eq.~\protect{\ref{ave}}),
   triangles: $m_i = 0$ (cartwheel rotation), 
   circles: $m_i = j_i$ (helicopter rotation). 
   The initial kinetic energy is $E_{i} = 10$~meV.}

\label{steric}
\end{figure}

Now in the system H$_2$/Pd(100) there are non-activated paths to dissociative
adsorption so that the concept of early and late barriers is not applicable.
Indeed fig.~\ref{h2pdstick} shows that by taking into account the rotational
population of the incoming beam in adsorption, the averaged sticking 
probability is slightly decreased as compared to molecules in 
the rotational ground state \cite{Gro95b} at all kinetic energies.
Especially at low energies additional rotational motion strongly reduces
the sticking probability. This is shown in fig.~\ref{steric} 
for a initial kinetic energy of $E_i = 10$~meV.
The diamonds correspond to the orientationally averaged sticking probability 
\begin{equation}\label{ave}
\bar S_{j_i} (E) \ = \ \frac{1}{2j_i +1} \ \sum_{m_i = -j_i}^{j_i} \ 
S_{j_i,m_i} (E). 
\end{equation}
The strong decrease is caused by a suppression of the steering effect. The
faster a molecule is rotating, the harder the molecular axis can be focused
to a favorable orientation towards adsorption, and the more molecules will
be reflected at the surface. 

In addition, fig.~\ref{steric} shows the effect of molecular orientation on
the sticking probability. The most favorable orientation to adsorption is
with the molecular axis parallel to the surface. Molecules with azimuthal
quantum number~$m = j$ have their axis preferentially oriented parallel to the 
surface. These molecules rotating in the so-called helicopter fashion  
dissociate more easily than molecules rotating in the cartwheel fashion 
($m = 0$) with their rotational axis preferentially parallel to the
surface  since the latter have a high probability hitting the
surface in an upright orientation in which they cannot dissociate. 
This steric effect has also been investigated in a number of model studies 
for purely activated adsorption \cite{Mow93,Dar94b,Bru94,Kum94,Dai95}.  

Invoking the principle of detailed balance, from the suppresion of the sticking
probability by rotation it follows that the population of
rotational states in desorption should be lower than expected for
molecules in thermal equilibrium with the surface temperature.
This so-called rotational cooling has indeed been found for H$_2$ 
molecules desorbing from Pd(100) \cite{Sch91} and is also well reproduced
by the six-dimensional quantum dynamical calculations \cite{Gro95b}.

\section{Conclusions}

In this brief review two recent examples of high-dimensional quantum dynamical
studies using {\em ab initio} potential energy surfaces have been presented.
For the interaction of light diatomic molecules interacting with metal
surfaces the continous excitation spectrum of the solid can be often 
neglected. For these systems now an almost complete understanding of the
dissociation process can be gained since translational, vibrational, rotational
and lateral effects no longer have to be treated separately. 
Our current understanding of simple surface reactions is mostly based
on low-dimensional dynamical studies on model potentials. These studies
provided us with a great deal of insight into the crucial principles
underlying these simple surface reactions.
However, high-dimensional studies do not only enhance the quantitative
agreement with experiment. Due to the microscopic information also a new
qualitative understanding of dissociation and scattering at surfaces 
can be obtained. Since this development has just started, 
many interesting results should be anticipated in the near future.

\section*{Acknowledgements}

It is a pleasure to acknowledge the many stimulating discussions and the
fruitful collaboration with Thomas Brunner, Bj{\o}rk~Hammer, Peter~Kratzer, 
Eckhard~Pehlke, Ralf~Russ, Steffen~Wilke, and other colleagues at the 
Technical 
University Munich and the Fritz-Haber-Institute, Berlin.  Special thanks go to
Wilhelm Brenig and Matthias Scheffler for guiding and supporting the
development of my work. I also want to thank George Darling and Stephen 
Holloway
for many discussions concerning dynamical matters and for organizing the 
Dynamics Workshop in Chester, U.K., which offered a great opportunity for
exchanging ideas on high-dimensional quantum dynamics.

\end{document}